# The influence of a Coulomb gap in the whole variable range hopping regime

J. F. Sampaio and A. Aparecido-Ferreira

*Departamento de Física, Universidade Federal de Minas Gerais, C.P. 702, Belo Horizonte, 31270-901, Brazil*

The temperature dependence of the electrical resistivity in insulator systems with a Coulomb gap in the density of states is expressed by a very simple function which coincides with the Efros-Shklovskii $T^{-1/2}$ result, at temperatures lower than some value $T_{\lim}$. Above this limit it consists of the product of the Mott $T^{-1/4}$ exponential with another one like a simply thermally activation process. It fits well several experimental results reported as having a crossover between the Efros-Shklovskii and Mott regimes, and allows the determination of the gap parameters even in experiments that do not reach the $T^{-1/2}$ regime. Also it accounts for some experimental results reported to follow a $T^{-\nu}$ behavior with $0.5 < \nu < 1$.

Hopping is a mechanism of electrical conduction by carriers in localized states [1,2]. It has been invoked to explain the electrical resistivity of the majority of different insulator materials. Examples are doped semiconductors, polymers, quasicrystals, DNA, granular systems, thin films, nanowires, carbon nanotubes, nanoparticle films, nanocrystals, and others. The underlying mechanism is that carriers jump from one localized state to another at a different place by thermally assisted tunneling. This implies a resistance $R \propto \exp(2r/\xi + \varepsilon/k_B T)$ between two sites, separated by a distance $r$, and with energies in the interval between $-\varepsilon$ and $\varepsilon$, relative to the Fermi level. Here $\xi$ is the localization length and $T$ the temperature. One can relate $r$ to $\varepsilon$ by requiring that in the spatial region $r^d$ ($d$ is the dimensionality of the system) there is one available state in the $\pm\varepsilon$ energy range. Thus defining $g(\varepsilon)$ as the density of states (DOS) per energy and per volume or area, in tri-dimensional (3D) or two-dimensional (2D) systems, respectively, one has [1],

$$N(\varepsilon) r^d = 1, \quad \text{with} \quad N(\varepsilon) \equiv \int_{-\varepsilon}^{\varepsilon} g(x) dx. \qquad (1)$$

By percolation arguments it is shown [1] that the resistivity $\rho$ has the same exponential behavior as $R$, with the inter site parameters $\varepsilon$ and $r$ replaced by some optimal hopping values $\varepsilon_h$ and $r_h$. These values are determined by the competition between the thermal and the tunneling processes. At temperatures not high enough to allow $r_h$ being the mean distance between next neighbors localized states, the carrier jump parameters $r_h$ and $\varepsilon_h$ will have a temperature dependence determined by the characteristic of the DOS at energies around the Fermi level. This is the variable range hopping (VRH) regime. A simple procedure which gives the right temperature dependence of these optimal parameters consists in minimizing $R$, with respect to $\varepsilon$ (or $r$). Using a constant DOS, $g(\varepsilon) \equiv g_o$, Mott [3] obtained a temperature depence for the given by

$$\ln[\rho(T)/\rho_o] = (T_o/T)^\nu, \qquad (2)$$

where $\nu = 1/(d+1)$ that equals 1/4 or 1/3 for 3D or 2D systems, respectively. Efros and Shklovskii (ES) [4] have shown that Coulomb interaction imposes the existence of a gap in the DOS which is parabolic, for 3D, and linear, for 2D systems. These forms can be expressed by $g(\varepsilon) \equiv g_o |\varepsilon/\Delta|^d$ and the parameter $\Delta$ is called the half-gap width. With this $g(\varepsilon)$ in Eq.(1), the minimization result of $R$ also has the form of Eq.(2), but with the exponent $\nu = 1/2$ for both $d$ values. Calling $T_M$ and $T_{ES}$ the constant $T_o$ for the Mott and the ES DOS, respectively, one has

$$k_B T_M = \beta_{M,d}/g_o \xi^d \; ; \; k_B T_{ES} = \beta_{ES,d} \left(\Delta^{d-1}/g_o \xi^d\right)^{1/d}. \qquad (3)$$

The constants $\beta_{M,d}$ and $\beta_{ES,d}$ have been determined by numerical percolation methods [1] because the described minimization procedure does not provide them correctly. There had been a intense discussion about the role of many-body effects on the resistivity because the $T^{-1/2}$ ES rule is based on a one-particle energy gap. But the conclusion is that the many-body effects also result in a $T^{-1/2}$ behavior with a little smaller $T_{ES}$ value [5].

Both, Mott and ES behaviors have been found in experimental results [1]. Biskupski *et al.* [6] have changed the resistivity of a sample from the Mott to the ES behavior by increasing the magnetic field. Zhang *et al.* [7] and Rosenbaum [8] have also obtained a crossover between a low temperature ES to a high temperature Mott behavior, as predicted by Efros and Shklovskii [4]. Massey and Lee [9] measured $\rho(T)$ in the crossover regime and also determined the Coulomb gap by tunneling spectroscopy. The results show that between the two regimes there is a crossover range of temperatures where none of them applies. There have been, however, some difficulties with the analysis of experimental data. There are results with a crossover temperature interval where both regimes seem to overlap [10]. Other, apparently, present the ES behavior at temperatures well above its temperature limit of validity, $T_{\lim}$ [7]. The $T_{\lim}$ definition itself is controversial, because different criteria have been used: $\varepsilon_h(\text{ES}) \approx \Delta$, $\varepsilon_h(\text{Mott}) \approx \varepsilon_h(\text{ES})$ or $k_B T_{\lim} \approx \Delta$. These problems cast doubts on the applicability of the VRH models. But, most probably, several experiments are in the temperature range of the crossover and the hopping parameters are not being correctly determined. There has been some effort [11, 12, 13, 14, 15, 16, 17] to find a formula able to describe the resistivity on the whole range of the VRH. This formula should result in the Mott behavior at high temperatures and at low ones in the ES or even in the stronger, $0.5 < s < 1$, behavior. Aharony et al. [11] have argued that the right side in Eq. (2) should have the scaling form $Af(T/T_x)$ where $f(x)$ is an universal function with the limiting behaviors $x^{-1/4}$ at $x \gg 1$ and $x^{-1/2}$ at $x \ll 1$. However the $f(x)$ they worked out fits well some results [18] and does not fit some other which seem to be related to hopping [10]. Their scaling form however has been proved in 3D by Meir [13] and for any dimension by Amir et al. [19]. In the references [12-17] the authors started already from a DOS formula which gives the ES gap [4] at low energies and tends to a constant, at high energies. None of them, however, leads to a simple formula for the resistivity. Besides, Nguyen and Rosenbaum [15, 16] obtained results seeming to follow Eq. (2), at low temperatures, but with $0.5 < \nu < 1$ which means a different energy dependence for the DOS function for each different $\nu$ value.

In this work we show that the truncated ES DOS functions shown in Fig. 1 (parabola for 3D and straight line for 2D) provide very simple analytical functions for the resistivity which have the Aharony et al. [11] scaling form and fit well the several experimental results used in the cited references. We express them by

$$g(\varepsilon) \equiv g_o |\varepsilon/\Delta|^{d-1}, \text{ for } |\varepsilon| \leq \Delta; \quad g(\varepsilon) \equiv g_o \text{ for } |\varepsilon| \geq \Delta. \quad (4)$$

From Eq. (1) one gets

$$N(\varepsilon) \equiv \begin{cases} 2\varepsilon g(\varepsilon)/d, & \varepsilon \leq \Delta \\ 2g_o[\varepsilon - \Delta(d-1)/d], & \varepsilon \geq \Delta \end{cases}. \quad (5)$$

Then we minimized, with respect to $\varepsilon$, the formula

$$\ln(\rho/\rho_o) = \alpha_1 2r(\varepsilon)/\xi + \alpha_2 \varepsilon/k_B T, \quad (6)$$

where $r(\varepsilon)$ was replaced by $N(\varepsilon)^{-1/d}$ and the constants $\alpha_1$ and $\alpha_2$ have been introduced with the purpose of adjusting the final results to those coming from percolation methods. This procedure resulted in

$$\ln(\rho/\rho_o) = \begin{cases} (T_{ES}/T)^{1/2}, & T \leq T_{\lim} \\ (T_M/T)^{1/(d+1)} + T_1/T, & T \geq T_{\lim} \end{cases}. \quad (7)$$

The parameters $T_M$ and $T_{ES}$ are the same given in Eq. (3); $T_1$, and $T_{\lim}$ are given by

$$T_1 = \frac{d^2-1}{4d^2} T_{ES} \left[\left(\frac{d+1}{2d}\right)^2 \frac{T_{ES}}{T_M}\right]^{1/(d-1)}, \quad (8)$$

$$k_B T_{\lim} = 4(\alpha_2 \Delta)^2 / k_B T_{ES}, \quad (9)$$

$$\alpha_2 \Delta = k_B T_1 d/(d-1). \quad (10)$$

The result in Eq. (7) is a continuous and differentiable function of $T$ having the exact ES behavior below the temperature $T_{\lim}$ above which another regime begins. This regime consists of the Mott function plus a simply thermally activated term. This term results from the fact that the evaluation of $N(\varepsilon)$ in Eq.(5), for $\varepsilon \geq \Delta$, considers the states in the gap. At high temperatures the Mott term can become dominant. So the $T^{-1}$ term makes the crossover between the ES and the Mott regimes. Because in the low and high temperature limits Eq. (7) gives the ES and the Mott regimes, the constants $\beta_{M,d}$ and $\beta_{ES,d}$ have to be the same already obtained from percolation methods.

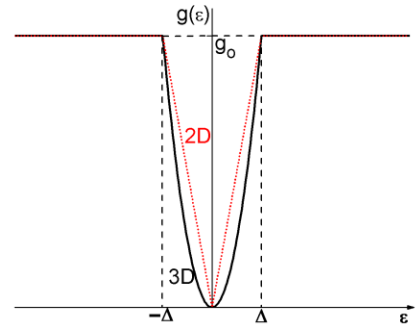

Fig. 1 – Density of states functions for 3D and 2D systems (truncated parabola and straight line).

This requires that $\alpha_1$ and $\alpha_2$ follow the relations $(d+1)^{d+1}(2\alpha_1)^{d-1} = 2^{3-1/d} d^{d+1/d} \beta_{M,d}/\beta_{ES,d}$ and $\alpha_2 = \beta_{ES,d}/[8(d/2)^{1/d}\alpha_1]$. With the values $\beta_{M,3} = 21.2$,

$\beta_{M,2} = 13.8$ and $\beta_{ES,3} = 2.8$ from the Ref. [1] and $\beta_{ES,2} = 6.2$ taken from Ref. [12] one has $\alpha_1 \approx 1.35$, $\alpha_2 \approx 0.226$, for 3D, and $\alpha_1 \approx 1.3$, $\alpha_2 \approx 0.59$, for 2D systems. One can see that here the ambiguity in the definition of $T_{\lim}$ disappears. It is the temperature where the optimal hopping energy, which is a continuous function of $T$, is equal to the gap half-width $\Delta$. Also the thermal energy $k_B T_{\lim}$ is much smaller than $\Delta$. This suggests that several experiments might have not reached $T_{\lim}$, and the procedure of evaluating $T_{ES}$, by fitting the low temperature data with the $T^{-1/2}$, becomes not reliable. And also, due to the crossover term $T_1/T$, the $T^{-1/4}$ fit of the high temperature data, in order to obtain $T_M$, requires a suitable high temperature range.

In order to test the applicability of Eq. (7) we show in Fig. 2 its fit to experimental results taken from the references [7, 9, 15, and 16]. The experimental data with the label 4 refers to a 2D sample and the other to 3D ones. The upper part of the figure shows $T^{-1/2}$ plots, and the lower part the $T^{-1/4}$ plots for 3D samples and $T^{-1/3}$ plot for the 2D sample. The fit parameters are shown in the upper part of Fig. 2. One can see that the fits are very good and this is also true for results, not included here, from other references. The vertical dotted lines in Fig. 2 correspond to the values of $T_{\lim}$ for each curve. Although the $T^{-1/2}$ plots seem to show good straight lines, none of the experimental results reached that limit, and the fitting procedure did not use the $T^{-1/2}$ part of Eq. (7). Even though, one could obtain from the fits both parameters $T_M$ and $T_{ES}$ and, consequently, the gap width.

From the slopes of the $T^{-1/2}$ and $T^{-1/4}$ plots of the data in curve 1 (Fig. 2) Zhang *et al.* [7] found the values $T_M = 51K$ and $T_{ES} = 2K$. These are comparable but a little bit smaller than the values we have obtained (Fig.2). They used for the the gap half-width $\Delta/k_B$ the formula $T_{ES}\sqrt{T_{ES}/T_M}$ which is the same we have when we use the $\alpha_2$ value obtained above. It results in a value equal to $0.40 K$ obtained from their parameters and $0.63 K$ from ours. For the data in curve 2, taken from Massey and Lee [9], one can get from the slopes of the $T^{-1/2}$ and $T^{-1/4}$ plots, $T_M = 831K$ and $T_{ES}$ between $11K$ to $15K$. The corresponding gap width is $\Delta \approx 0.14 \text{meV}$, also lower than the value $\Delta \approx 0.22 \text{meV}$ one obtains from our fitting parameters shown in Fig. 2.

The experimental results in the curves 3 and 4 of Fig. 2 are, respectively from a 3D and a 2D film of $Ni_xSi_{1-x}$, and have been taken from Nguyen and Rosenbaum [16, 15], who fitted them with a set of four equations originated from a DOS function with a dependence on $\varepsilon^n$. It determines a low temperature behavior like the one given in Eq. (2) but with an exponent $\nu$ assuming different values according to the $n$ value. They found $\nu = 0.72$ for the curve 3 and $\nu = 0.77$ for the curve 4. Their fit curves are included in Fig. 2, by dotted lines (blue color online), which practically coincide with ours curves. We could fit both curves with Eq. (7) whose combination between the Mott and the $T^{-1}$ terms explains well the $\nu > 1/2$ trends found by the authors. The fit with Eq. (7), whose parameters are shown in Fig. 2, lead to values for the half-gap widths which, coincidently, are about the same for both samples, $\Delta/k_B \approx 26K$.

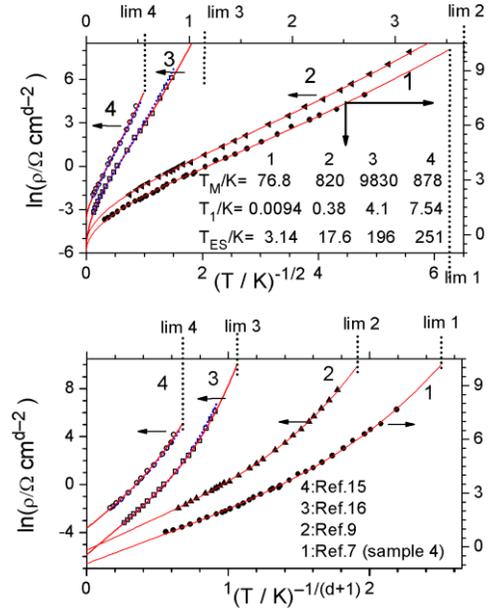

Fig.2 – Fitting curves (solid lines) of experimental results for $\ln(\rho)$ (plots with symbols) taken from four different sources listed in the lower part. The horizontal axis is $T^{-1/2}$ in the upper part which shows also the fitting parameters. In the lower part it is $T^{-1/3}$ for the curve 4 (2D system) and $T^{-1/4}$ for the other curves.

One can express Eq. (7) in the scaling format $\ln(\rho/\rho_o) = A f(x)$ prescribed by Aharony *et al.* [11]. We find it more convenient to define $x \equiv T_{\lim}/T$ because $T$ appears always in the denominator. Defining also $A \equiv (T_{ES}/T_{\lim})^{1/2}$, one has

$$f_d(x) = \begin{cases} x^{1/2}, & x \geq 1 \\ \dfrac{d+1}{2d}\dfrac{1}{x^{d+1}} + \dfrac{d-1}{2d}x, & x \leq 1 \end{cases}. \quad (11)$$

The study of this function is very helpful in order to analyze appropriately experimental results. In Fig. 3 we show plots of $f_3(x)$ versus $x^{1/4}$ (curve 1) and $x^{1/2}$ (curve 2). The second one shows that, although the ES regime happens only at $x \geq 1$, $f_3(x)$ matches the $x^{1/2}$ (ES) straight line for $\sqrt{x} \geq 0.84$. So a $T^{-1/2}$ fit will result in a good value for the $T_{ES}$ parameter for $T \leq 1.4 T_{lim}$. This is also true for 2D systems. For temperatures above this value one can see that the slopes of the curve 2 will be smaller than that of the straight line, and the $T_{ES}$ values obtained from the slope will be underestimated. However in the region $\sqrt{x} \leq 0.134$ of curve 2, meaning temperatures in the range $T \geq 56 T_{lim}$, the curve's slopes become higher than that of the ES rule. For 2D systems this happens for $T \geq 39 T_{lim}$. The curve 1 of Fig. 3 matches approximately the $x^{1/4}$ straight line (Mott regime) for $x^{1/4} \leq 0.33$, i.e., $T \geq 80 T_{lim}$. This defines the temperature range $1.4 T_{lim} < T < 80 T_{lim}$ for the crossover from the Mott to the ES regime. The crossover range for 2D systems is $1.4 T_{lim} < T < 37 T_{lim}$. When we fit experimental data with $T^{-1/4}$ or $T^{-1/3}$ in order to obtain the Mott parameter $T_M$, at temperatures above these ranges, the obtained value for $T_M$ will be overestimated. This explains the $T_M$ value we have obtained for the data in curve 4 to be about one half the value obtained by their authors [15]. For these data we obtained $T_{lim} \approx 3.6 K$ meaning that the Mott regime extends to $134 K$. The Mott fit in Ref. [15], however, goes down to $25 K$. The inset in Fig. 3 shows a portion of the curve $f_3(x)$ well fitted by $a + bx^\nu$, with $\nu = 0.76$. A different portion will be fitted with a different $\nu$ value. This shows that, depending on the sample and on the temperature interval of experiments, the resistivity results may be described by some part of the function in Eq. (7), seeming to follow Eq. (2) with some $\nu$ value above 0.5. This what happens to the results in Refs. [16,15].

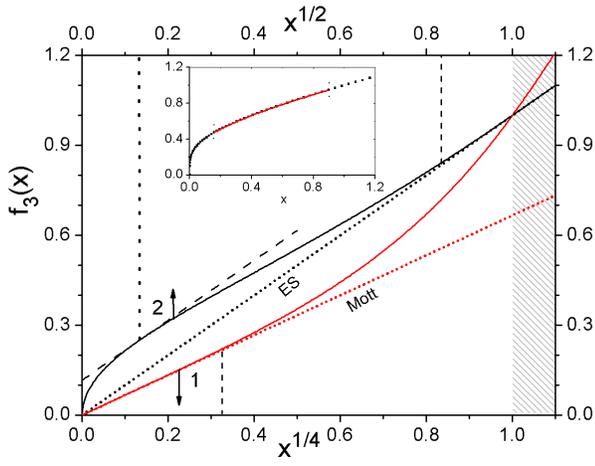

Fig. 3: Plots of $f_3(x)$ versus $x^{1/4}$ (curve 1, lower scale) and $x^{1/2}$ (curve 2, upper scale). The straight lines going out from the origin represent the Mott and ES models. INSET: $f_3(x)$ versus $x$ (dots) with a part fitted by $a + bx^\nu$ (solid line).

We conclude by saying that the DOS function we used here is a good effective replacement for other more realistic functions and results in a much simpler resistivity formula. It fits experimental data on the whole VRH regime and allows the easy obtainment of the effective gap parameters. It can also explain well the experimental results whose temperature dependence seems to require $0.5 < \nu < 1$. As for this it is important to realize that if we want a $\ln(\rho/\rho_o)$ in the scaling form proposed by Aharony et al. [11], the values of $\ln(\rho_o)$ obtained from the extrapolation of the low and high temperatures experimental data by a $T^{-\nu}$ formula, should have the same value. This work shows also that, probably, several experimental results may not have achieved the true ES regime, and the DOS parameters have not been well determined.

ACKNOWLEDGMENTS – We are grateful to H. D. Pfannes for reading and commenting on this work. We acknowledge the brazilian agencies CNPq and FAPEMIG for financial support and CAPES for providing access to scientific journals.